\documentclass[]{rmf-d}
\usepackage{nopageno,rmfbib,multicol,times,epsf,amsmath,amssymb}
\usepackage[latin1]{inputenc}
\usepackage[spanish,english]{babel}
\usepackage[]{caption2}
\newcommand{\adiciona}{\addtocounter{equation}{1}}
\newcommand{\subeq}[1]{\tag{\theequation$\,#1$}}
\newcommand{\del}{\delta^{2}(\vec{x} - \vec{y})}
\def\rmfcornisa{ \hfill\rmf\ \hfill }
\def\rmfcintilla{{\it Rev.\ Mex.\ F\'{\i}s.\/}}

\clearpage
\rmfcaptionstyle
\begin{document}

\title{Covariancia de la teoría Autodual Vectorial}
\author{Pío J. Arias $^{1,2}$ y Mónica A. García-Ñustes$^{1}$\footnote{Poster presentado en el IV Congreso de S.V.F}}
\address{$^{1}$Centro de Física Teórica y Computacional,Facultad de Ciencias, Universidad Central de Venezuela, AP 47270, Caracas
1041-A, Venezuela}
\address{$^{2}$Centro de Astro\'isica Te\'orica, Facultad de Ciencias ULA, La Hechicera, M\'erida 5101, Venezuela}

\maketitle
\begin{abstract}
The Poisson algebra between the fields involved in the vectorial selfdual action is obtained by means of the reduced action. The conserved charges associated with the invariance under the inhomogeneous Lorentz  group are obtained and its action on the fields. The covariance of the theory is proved using the Schwinger-Dirac algebra. The spin of the excitations is discussed.
\end{abstract}
\keys{Lagrangian formulation, Poincar\'e algebra}


\begin{resumen}
A partir de la acción reducida de la teoría autodual vectorial en 2+1 se obtiene el \'algebra de Poisson entre los campos que intervienen
en la acci\'on original. Las cargas conservadas asociadas a la invariancia bajo el grupo inhomog\'eneo de Lorentz son calculados  así,
como su acci\'on sobre los campos. El \'algebra de Schwinger es obtenida y con ella se muestra la covariancia de la teo\'ia. Se discute 
el spin de las excitaciones.
\end{resumen}

\vspace{-2pt} \descript{Formulaci\'on Lagrangiana,
\'algebra de Poincar\'e}

\pacs{$11.10.Ef$, $03.70.+k$, $11.30.Cp$}

\begin{multicols}{2}
El estudio de teorías vectoriales y tensoriales en  2+1 dimensiones
estuvo inicialmente motivado por su conexi\'on con el
comportamiento de modelos, en 3+1 dimensiones, a altas
temperaturas$^{1}$. Sin embargo, en la actualidad, la física planar posee un inter\'es real intrínseco, ya que presenta
características propias que hacen sumamente atractivo
su estudio y análisis en el contexto de la teoría cuántica de campos. En el presente trabajo analizamos
la teoría autodual vectorial masiva en 2+1 dimensiones, obteni\'endose su espectro f\'isico, 
se obtienen los generadores de las transformaciones infinitesimales y  analizamos su covariancia. La covariancia de la teoría es
mostrada usando el \'Algebra de Schwinger-Dirac.

\vspace{1pt}

La teoría autodual vectorial esta descrita por la acción$^{2}$
\begin{equation}
  S = - \frac{\mu}{2}\int \,d^{3}x \,\left(\varepsilon^{{\mu}{\nu}{\lambda}} A_{\mu} \partial_{\nu} A_{\lambda}
+ \mu A_{\mu}A^{\mu}\right).
\end{equation}
Esta acción no posee invariancia locales y puede mostrarse que corresponde a una fijaci\'on de calibre de la teor\'ia topol\'ogica
masiva vectorial$^{{3},{4},{5},{6},{7}}$. Adem\'as, ambos modelos puede probarse que son duales uno del otro$^{{7},{8},{9}}$.
Las ecuaciones de  movimiento, son:
\begin{equation}
\label{ec:eje0}
 \varepsilon^{{\mu}{\nu}{\lambda}} \partial_{\nu} A_{\lambda} + \mu A^{\mu} = 0,
\end{equation}
de donde $A_{o} = \frac{1}{\mu} \varepsilon_{{i}{j}} \partial_{i} A_{j}$ 
con $\varepsilon_{{i}{j}} = \varepsilon^{{o}{i}{j}}$.

Si hacemos la descomposición 2+1 de la acción,
utilizando la metrica $\eta_{{\mu}{\nu}} = (- + +)$, obtenemos
\begin{equation}
\begin{split}
S &= - \frac{\mu}{2}\int d^{3}x \,\left(\varepsilon^{{o}{i}{j}} \left(A_{o} \partial_{i} A_{j} - A_{i} \dot{A^{j}} + A_{i} \partial_{j} A_{o} \right) \right)\\
&\quad + \int d^{3}x \left( \mu A_{i}A^{i} - \mu A_{o}A_{o}  \right),
\end{split}
\end{equation}
de donde utilizando la descomposición transverso-longitudinal
$$A_{i}  =  \varepsilon_{{i}{j}}\partial_{j} A_{T} + \partial_{i} A_{L} \equiv A_{i}^{T} + A_{i}^{L}, $$
con $A_{i}^{T}$  es la parte transversal de $A_{i}$ y $A_{i}^{L}$  es la parte longitudinal de $A_{i},$
y junto con la  expresi\'on (\ref{ec:eje1}) llegamos a,
\begin{equation}
\begin{split}
S &= \int d^{3}x \left( - \frac{1}{2}(-\Delta) A_{T}(-\Delta) A_{T} - \frac{{\mu}²}{2} A_{T}(-\Delta) A_{T}\right)\\
& - \int d^{3}x \, \left( \frac{{\mu}²}{2} A_{L} (-\Delta) A_{L} - \, \mu \dot{A_{T}}(-\Delta) A_{L} \right),
\end{split}
\end{equation}
donde $\Delta = \partial_{i}\partial_{i}$.

Si hacemos las definiciones,
\adiciona
\begin{align}
\label{ec:eje10}
Q &= (-\Delta)^{1/2} A_{T} \subeq{a}\\
\label{ec:eje12}
\Pi &= - \, \mu (-\Delta)^{1/2}  A_{L} , \subeq{b}
\end{align}
podemos llegar a la acción reducida,
\begin{equation}
\label{ec:eje5}
S =  \int d^{3}x \, \left(\dot{Q} \, \Pi - \frac{1}{2} Q(-\Delta +
\mu^2) Q - \frac{1}{2} \Pi \,\Pi \right),
\end{equation}
que corresponde a la acci\'on de una excitac\'on de masa $\mu$ y con Hamiltoniano  definido positivo.
\newline
La ventaja de (\ref{ec:eje5}) es que permite obtener directamente los corchetes de Poisson entre las variables originales.
Postulando los corchetes fundamentales entre el campo $Q(x)$ y su momentos conjugado $\Pi(x)$, 
usando (\ref{ec:eje1}) y (6) obtenemos,
\adiciona
\begin{align}
\label{ec:eje8}
\left\{A_{o}(x) \, , A_{o}(y) \right\} &= 0 \subeq{a}\\
\left\{A_{o}(x) \, , A_{i}(y) \right\} &= \frac{1}{{\mu}^2}\partial_{i}\del \subeq{b} \\
\left\{A_{i}(x) \, , A_{j}(y) \right\} &= -\frac{1}{\mu}\varepsilon_{{i}{j}}\del \subeq{c}
\end{align}
los cuales son consistentes con (\ref{ec:eje1}).

Veamos, ahora, las cantidades conservadas asociadas a invariancias de la teor\'ia bajo el grupo inhomog\'eneo de Lorentz.
Asociado a las traslaciones tenemos el tensor de energía-momento canónico de la teoría,
\begin{equation}
T^{\mu}{}_{\nu} = \mbox{\textit{L}}\delta^{\mu}{}_{\nu} - \frac{\partial \mbox{\textit{L}}}{\partial(\partial_{\mu}A_{\rho})} \partial_{\nu}A_{\rho}
\end{equation}
\begin{equation}
T^{\mu}{}_{\nu} = - \frac{\mu}{2} \varepsilon^{{\beta}{\mu}{\sigma}} \left(A_{\nu}\partial_{\sigma}A_{\beta} - A_{\beta}\partial_{\sigma}A_{\nu}\right) - \frac{{\mu}²}{2} A^{\nu}A_{\nu}\delta^{\mu}{}_{\nu}
\end{equation}
este satisface $\partial_{\mu} T^{{\mu}{\nu}} = 0$ sobre (\ref{ec:eje0}).
De igual forma, asociadas a las rotaciones espacio-temporales tenemos el tensor de densidad
de momento angular,
\begin{equation}
M^{{\mu}{\alpha}{\beta}} = -  \varepsilon^{{\alpha}{\beta}{\rho}} \left[ \varepsilon_{{\rho}{\sigma}{\mu}}T^{{\mu}{\sigma}}x^{\lambda}  + \frac{\partial \mbox{\textit{L}}}{\partial(\partial_{\mu}A_{\theta})} \varepsilon_{{\rho}{\theta}}{}{}^{\nu}A_{\nu}\right]
\end{equation}
el cual satisface $\partial_{\mu} M^{{\mu}{\alpha}{\beta}} = 0$ sobre (\ref{ec:eje0}).
\newline
Se sabe que a partir de $T^{\mu}{}_{\nu}$, podemos construir el
tensor simétrico, de Belifante, $\Theta^{{\mu}{\nu}}_{B}$$^{10}$, el cual resulta ser ``on shell''.
\begin{equation}
\label{ec:eje15}
\Theta^{{\mu}{\nu}}_{B} =  m^{2} A^{\mu} A^{\nu} - \frac{{m}²}{2} \eta^{{\mu}{\nu}}A^{\alpha} A_{\alpha}
\end{equation}
$\Theta^{{\mu}{\nu}}_{B} $ es conservado y sus cargas asociadas son las mismas
de $T^{{\mu}{\nu}}$.Además, podemos construir un  tensor equivalente al tensor de densidad de momento angular $M^{{\mu}{\alpha}{\beta}}$ 
utilizando el tensor  $\Theta^{{\mu}{\nu}}_{B}$,
\begin{equation}
\label{ec:eje16}
 M^{{\mu}{\alpha}{\beta}}_{B} = x^{\beta} \Theta^{{\mu}{\alpha}}_{B} - x^{\alpha}\Theta^{{\mu}{\beta}}_{B}
\end{equation}
$M^{{\mu}{\alpha}{\beta}}_{B}$ es trivialmente conservado debido a la conservaci\'on de $\Theta^{{\mu}{\nu}}_{B}$ y 
sus cargas asociadas son las mismas que $M^{{\mu}{\alpha}{\beta}}$.

Las cargas conservadas asociadas al $\Theta^{{\mu}{\nu}}_{B}$ y al $M^{{\mu}{\alpha}{\beta}}_{B}$ son,
\begin{equation}
\begin{split}
\label{ec:eje17}
\mbox{P}^{\mu}_{B} &= \int d^2x \Theta^{{o}{\mu}}_{B} \equiv \int \, d^2x \, \mathcal{P}^{\mu}\\
&= \int d^2x \left(m^{2} A^{o} A^{\nu} - \frac{{m}²}{2} \eta^{{o}{\mu}}A^{\alpha} A_{\alpha}\right),
\end{split}
\end{equation}
y
\begin{equation}
\begin{split}
\label{ec:eje18}
\mbox{{\large J}}^{{\alpha}{\beta}}_{B} &= \int d²x \left(\Theta^{{o}{\mu}}_{B} x^{\nu} - \Theta^{{o}{\nu}}_{B}x^{\mu} \right) \\
&= - \int d²x \varepsilon^{{\alpha}{\beta}{\rho}} \varepsilon_{{\rho}{\sigma}{\lambda}} \mathcal{P}^{\sigma}_{B}x^{\lambda}
\end{split}
\end{equation}
\adiciona
respectivamente. Ahora deseamos comprobar que \'estas efectivamente generan las transformaciones de los campos. Para esto calculamos,
los corchetes de Poisson, usando (10) y (\ref{ec:eje17}), 
\adiciona
\begin{align}
\label{ec:eje19}
&\left\{A_{o}(x)\,,\varepsilon\,P^{o}_{B}(y)\right\} =  \varepsilon \partial_{i} A_{i} \subeq{a} \\
&\left\{A_{o}(x)\,,\varepsilon_{i}\,P^{i}_{B}(y)\right\} = - \varepsilon_{i}\partial_{i}A_{o} \subeq{b} \\
&\left\{A_{i}(x)\,,\varepsilon\,P^{o}_{B}(y)\right\} = \varepsilon \left(\partial_{i}A_{o}
 - \mu \, \varepsilon_{{i}{j}}A_{j}\right) \subeq{c} \\
&\left\{A_{i}(x)\,,\,\varepsilon_{j}\,P^{j}_{B}(y)\right\} = -\varepsilon_{j}\partial_{j}A_{i}  \subeq{d}
\end{align}
usando las ecuaciones de movimiento (\ref{ec:eje0}) es inmediato ver que,
\begin{equation*}
\left\{A_{\mu}, \varepsilon_{\nu}P^{\nu}_{B} \right\} = - \varepsilon_{\nu}\partial^{\nu}A_{\mu}
\end{equation*}
con $\varepsilon_{\mu}=(\varepsilon,\varepsilon_{j})$.
An\'alogamente calculamos con (10) y (\ref{ec:eje18})
\begin{equation}
\begin{split}
&\left\{A_{i}(x)\,,\frac{1}{2}\omega_{{\alpha}{\beta}} \mbox{{\Large J}}^{{\alpha}{\beta}}_{B}(y)\right\} = \omega_{{o}{l}} x^{o}
\left(\partial_{i}A_{l} - \mu \varepsilon_{{i}{l}} A_{o}\right) \\
& \qquad + \omega_{{o}{l}}x^{l}\left(\partial_{i}A_{o} - \mu \varepsilon_{{i}{k}}A_{k} + \delta_{i}{}^{l}A_{o}\right)\\
& \qquad + \frac{1}{2}\omega_{{k}{l}}\varepsilon_{{k}{l}}\left( \varepsilon_{{i}{j}}A_{j} - \mu x^{i} A_{o} - \varepsilon_{{m}{n}}x^{n}
\partial_{i} A_{m}\right)
 \end{split}
\end{equation}
\begin{equation}
\begin{split}
&\left\{A_{o}(x)\,,\frac{1}{2}\omega_{{\alpha}{\beta}} \mbox{{\Large J}}^{{\alpha}{\beta}}_{B}(y)\right\} =
 \frac{1}{2}\omega_{{k}{l}}\varepsilon_{{k}{l}} \varepsilon_{{i}{j}} x^{i} \partial_{j} A_{o}\\
& \qquad + \omega_{{o}{l}}\left( A_{l} + x^{o}\partial_{l}A_{o} +  x^{l} \partial_{k}A_{k}\right)
\end{split}
\end{equation}
que con (\ref{ec:eje0}) nos lleva a,
\begin{equation}
\begin{split}
\left\{A_{\mu}(x)\,,\frac{1}{2}\omega_{{\alpha}{\beta}} \mbox{{\Large J}}^{{\alpha}{\beta}}_{B}(y)\right\} = \\
 - \frac{1}{2}\omega_{{\delta}{\theta}}\varepsilon^{{\delta}{\theta}{\lambda}}\left[\varepsilon_{{\lambda}{\mu}}{}{}^{\nu}A_{\nu} + 
\varepsilon_{{\lambda}{\rho}}{}{}^{\nu}x^{\rho}\partial_{\nu}A_{\mu}\right]
\end{split}
\end{equation}
que corresponde a la transformaci\'on infinitesimal del campo $A_{\mu}$ para una transformaci\'on de Lorentz. 
Para una transformaci\'on combinada tendremos entonces que
\begin{equation}
\delta A_{\mu} = \left\{ A_{\mu}\,,\, - \varepsilon_{\nu}P^{\nu} + \frac{1}{2} \omega_{{\alpha}{\beta}}\mbox{{\Large J}}^{{\alpha}{\beta}} \right\}.
\end{equation}
Prosiguiendo con nuestro an\'alisis, deseamos comprobar que la teoría es invariante relativista para esto observamos  (\ref{ec:eje1}) y
utilizando el reemplazo $\{\quad\} \to -i[\quad],$
tenemos,
\begin{align}
[\Theta^{{o}{o}}(x),\Theta^{{o}{o}}(y)] &= - i\left( \Theta^{{o}{i}}(x) + \Theta^{{o}{i}}(y) \right)\partial_{i}\del \subeq{a} \\
[\Theta^{{o}{o}}(x),\Theta^{{o}{i}}(y)] &= - i\left( \Theta^{{i}{j}}(x) + \Theta^{{o}{o}}(y)\delta^{{i}{j}} \right)\partial_{j}\del \subeq{b} \\
[\Theta^{{o}{i}}(x),\Theta^{{o}{j}}(y)] &= - i\left( \Theta^{{o}{j}}(x)\partial_{i} + \Theta^{{o}{i}}(y)\partial_{j} \right)\del. \subeq{c}
\end{align}
que es justamente el álgebra de Schwinger-Dirac, para una teoría vestorial masiva. \'Esta garantiza que las cargas conservadas asociadas
$\mbox{\textrm{P}}^{\mu}$ y $\mbox{{\large{J}}}^{{\alpha}{\beta}}$ obedezcan el álgebra de Poincaré $^{11}$. Para ver esto, notamos que
\begin{equation*}
\begin{split}
&\left [\mbox{\textrm{P}}^{o}\,,\,\Theta^{{o}{i}}(x)\right] 
=\int d^{2}y [\Theta^{{o}{o}}(y),\Theta^{{o}{i}}(x)]\\
&= \int d^{2}y \left(\Theta^{{i}{j}}(y) + \eta^{{i}{j}}\Theta^{{o}{o}}(x)\right) \partial_{j}{}^{y}\del \\
&= - i \partial_{o}{}^{x}\Theta^{{i}{o}}(x)\\
\end{split}
\end{equation*}
como $\dot{\mbox{\textrm{P}}^{\mu}} = 0$ entonces $\left [\mbox{\textrm{P}}^{o}\,,\,\Theta^{{o}{i}}\right] = 0$
de donde $[\mbox{\textrm{P}}^{0},\mbox{\textrm{P}}^{j}] = 0$.

As\'i mismo,
\begin{equation*}
\begin{split}
&\left [\mbox{{\large{J}}}^{{o}{i}}\,,\,\mbox{\textrm{P}}^{o}\right] = \int d^{2}y \,d^2x [\Theta^{{o}{o}}(x)x^{i} - \Theta^{{o}{i}}(x)t
\,,\Theta^{{o}{o}}(y)]\\
&= \int d^{2}y  x^{i}\left[\Theta^{{o}{o}}(x), \mbox{\textrm{P}}^{o}\right] - t\left[\Theta^{{o}{i}}(x), \mbox{\textrm{P}}^{o}\right] \\
&= - i\int d^{2}x x^{i}\partial_{j}{}^{x}\Theta^{{o}{j}} = i \int d^{2}x\Theta^{{o}{i}} = i \mbox{\textrm{P}}^{i}
\end{split}
\end{equation*}

Procediendo de manera an\'aloga obtenemos
\begin{equation*}
\begin{split}
&[\mbox{{\large{J}}}^{{i}{j}},\mbox{\textrm{P}}^{k}] =  i\left( \delta^{j}_{k}\mbox{\textrm{P}}^{i} - \delta^{i}_{k}
\mbox{\textrm{P}}{j} \right), \,\, 
[\mbox{{\large{J}}}^{{i}{j}},\mbox{\textrm{P}}^{o}] = 0  \\
& \qquad \quad \qquad[\mbox{{\large{J}}}^{{o}{i}},\mbox{\textrm{P}}^{j}] =  i\delta^{{i}{j}}\mbox{\textrm{P}}^{o}
\end{split}
\end{equation*}
\begin{align*}
&[\mbox{{\large{J}}}^{{o}{i}},\mbox{{\large{J}}}^{{o}{j}}] =  i\mbox{{\large{J}}}^{{i}{j}}, \,\,
[\mbox{{\large{J}}}^{{o}{i}},\mbox{{\large{J}}}^{{k}{l}}] =  i\left(\mbox{{\large{J}}}^{{o}{l}}\eta^{{i}{k}} -
\mbox{{\large{J}}}^{{o}{k}}\eta^{{i}{l}}\right)\\
&[\mbox{{\large{J}}}^{{i}{k}},\mbox{{\large{J}}}^{{m}{l}}] =  i\left(\mbox{{\large{J}}}^{{i}{l}}\eta^{{k}{m}} + 
\mbox{{\large{J}}}^{{l}{k}}\eta^{{i}{m}} - \mbox{{\large{J}}}^{{m}{k}}\eta^{{i}{l}} - \mbox{{\large{J}}}^{{i}{m}}\eta^{{k}{l}}\right)
\end{align*}

Todos los corchetes se escriben as\'i de manera covariante como
\begin{equation}
[\mbox{\textrm{P}}^{\mu},\mbox{\textrm{P}}^{\nu}] = 0
\end{equation}
\begin{equation}
[\mbox{{\large{J}}}^{{\mu}{\nu}},\mbox{\textrm{P}}^{\lambda}] =  i\left( \mbox{\textrm{P}}^{\mu} \eta^{{\nu}{\lambda}} -  \mbox{\textrm{P}}^{\nu} \eta^{{\mu}{\lambda}}\right)
\end{equation}
\begin{equation}
[\mbox{{\large{J}}}^{{\mu}{\nu}},\mbox{{\large{J}}}^{{\sigma}{\rho}}] =  i\left(\mbox{{\large{J}}}^{{\nu}{\sigma}}\eta^{{\mu}{\rho}}
 + \mbox{{\large{J}}}^{{\mu}{\rho}}\eta^{{\nu}{\sigma}} - \mbox{{\large{J}}}^{{\nu}{\rho}}\eta^{{\mu}{\sigma}} - \mbox{{\large{J}}}^{{\mu}
{\sigma}}\eta^{{\nu}{\rho}}\right)
\end{equation}
esto \'ultimo es el \'algebra  de ISO(2,1) en 2+1 dimensiones, quedando as\'i mostrada la invariancia de la teoría Autodual 
Vectorial bajo transformaciones del grupo de Poincaré.

Por último analicemos la contribución del spin en la teoría realizando el siguiente procedimiento sobre el espacio de momento.
Haciendo la siguiente expansión en modos para Q(x)$^{3}$, 
\begin{equation}
\mbox{Q}(x) = \int \frac{d^{2}\vec{k}}{2\pi\sqrt{2\omega(\vec{k})}}\left[ e^{ik.x}a(\vec{k}) + e^{-ik.x}a^{\dag}(\vec{k}) \right],
\end{equation}
donde $k^{\mu}=(\omega(\vec{k}),\vec{k})$, con $\omega(\vec{k})=\sqrt{\vec{k}^{2}+\mu^{2}}$ y 
$[a(\vec{k}),a^{\dag}(\vec{k'})]=\delta(\vec{k}-\vec{k'})$. Encontramos que 
los generadores son como los del campo escalar, excepto para los boosts de Lorentz, en donde se
encuentra un termino extra, el cual introduce lo que parece una anomal\'ia a la
teor\'ia
\begin{equation}
\begin{split}
\mbox{{\large{J}}}^{{o}{i}} &= \frac{i}{2} \int\, d^2\vec{k}\,\omega(\vec{k}) [a^{\dag}(\vec{k}) \overleftrightarrow{\partial_{i}} 
a(\vec{k})] \\
& - \mu\int d^{2}\vec{k}\varepsilon^{{i}{j}}\frac{k^{i}}{\vec{k}^{2}}a^{\dag}(\vec{k})a(\vec{k})
\end{split}
\end{equation}
Esto proviene del hecho de que el campo que describe la única excitación posible, no es realmente un escalar. 
A fin de evitar esta ``anomal\'ia'' se redefine la fase de los operadores de creación y  aniquilación de 
la forma $a(\vec{k}) \rightarrow e^{i(\mu/\mid\mu\mid)\theta}a(\vec{k})$. De esta forma el generador de 
rotaciones adquiere un termino de spin,
\begin{equation}
\mbox{{\large{J}}} = \int\, d^2\vec{k}\,a^{\dag}(\vec{k})\frac{1}{i}\frac{\partial}{\partial\theta}a(\vec{k}) + \frac{\mu}{\mid\mu\mid}
\int d^{2}\vec{k} a^{\dag}(\vec{k})a(\vec{k}),
\end{equation}
mientras que los generadores de los boosts toman la forma,
\begin{equation}
\begin{split}
\mbox{{\large{J}}}^{{o}{i}} &= \frac{i}{2} \int\, d^2\vec{k}\,[a^{\dag}(\vec{k}) \overleftrightarrow{\partial_{i}} a(\vec{k})] \\
& + \frac{\mu}{\mid\mu\mid}\int d^{2}\vec{k} \frac{1}{\omega(\vec{k})+\mid\mu\mid}\varepsilon^{{i}{j}}
k^{j}a^{\dag}(\vec{k})a(\vec{k})
\end{split}
\end{equation}
en el cual la ``anomal\'ia''  ya no se encuentra.

Lo anterior pone en evidencia un spin $\mu/\mid\mu\mid$, cuyo signo depende del termino de 
Chern-Simons de la acción, el cual no afecta la positividad del Hamiltoniano. Este resultado se obtiene exactamente como con el caso de la teor\'ia 
topol\'ogica masiva$^{3}$.  

Para concluir hemos observado que la teor\'ia autodual vactorial es covariante de Poincar\'e y que las cargas asociadas a  la 
invariancia bajo este grupo son las generadoras de las transformaciones infinitesimales como era de esperarse. La discusi\'on en 
relaci\'on al spin de las excitaciones se da de igual manera que en la teor\'ia topol\'ogica masiva. Toda la discusi\'on parti\'o de 
llevar la acci\'on a sus grados f\'isicos de libertad sin necesidad de hacer el procedimiento de Dirac para lagrangianos singulares. 
  
\vspace{1pt}
{\bf Agradecimientos}
P.J.A. agradece el apoyo recibido por parte de la OPSU. Este trabajo es parcialmente apoyado por el
proyecto G-2001000712 del FONACIT.

\end{multicols}

\medline

\begin{multicols}{2}


\end{multicols}
\end{document}